\documentclass[12pt]{article}
\usepackage{amsmath,amssymb}
\newcommand{\beq}{\begin{equation}}
\newcommand{\eeq}{\end{equation}}
\newcommand{\ba}{\begin{array}}
\newcommand{\ea}{\end{array}}
\newcommand{\bea}{\begin{eqnarray}}
\newcommand{\eea}{\end{eqnarray}}
\newcommand{\bean}{\begin{eqnarray*}}
\newcommand{\eean}{\end{eqnarray*}}

\newtheorem{theorem}{Theorem}[section]
\newtheorem{prop}[theorem]{Proposition}
\newtheorem{lemma}{Lemma}
\newtheorem{defi}[theorem]{Definition}

\newenvironment{pf}{\begin{proof} \rm}{\hfill$\square$ \end{proof}}
\newcounter{appendix}
\newcommand{\newappendix}[1]{\vspace{2mm}\pagebreak[3]
\addtocounter{appendix}{1}
\renewcommand{\theequation}{\Alph{appendix}.\arabic{equation}}
\setcounter{equation}{0}
\begin{flushleft}{\Large\bf Appendix \Alph{appendix} #1}
\end{flushleft}\nopagebreak\medskip\nopagebreak}

\newcommand{\CD}{{\cal D}}

\newcommand{\CZ}{{\cal Z}}
\newcommand{\CS}{{\cal S}}

\newcommand{\CM}{{ M}}

\newcommand{\CL}{{\cal L}}

\makeatletter
\@addtoreset{equation}{section}
\renewcommand{\theequation}{\thesection.\arabic{equation}}
\makeatother

\newcommand{\DD}{{\mathbb D}}

\def\be{\beta}
\def\al{\alpha}

\def\la{\lambda}

\newcommand{\cmp}[3]{Comm. Math. Phys. {\bf #1} (#2), #3}

\newcommand{\lmp}[3]{Lett. Math. Phys. {\bf #1} (#2), #3}

\newcommand{\jmp}[3]{Jour. Math. Phys. {\bf #1} (#2), #3}
\newcommand{\rref}[1]{(\ref{#1})} 

\def\dsl{\displaystyle}

\newcommand{\wid}[1]{{\widetilde{#1}}}

\def\mat2#1#2#3#4{{\left(\begin{array}{cc}#1 & #2\\ #3 & #4
      \end{array}\right)}}

\def\mats2#1#2#3#4{{\left(\begin{array}{cc}#1 & #2\vspace{2truemm} \\ #3 & #4 
\end{array}\right)}}

\def\ddd#1#2{\displaystyle{\frac{\partial #1}{\partial #2}}}

\newcommand{\Ha}[1]{H^{(#1)}}
\newcommand{\Ka}[1]{K^{(#1)}}

\def\Nij{Nijenhuis}

\def\endpf{\begin{flushright}$\square$\end{flushright}}

\def\alg{{\mathfrak g}}
\def\alp{{\mathfrak p}_-}

\def\ger{hierarch}
\def\var{manifold}

\def\bih{biham\-il\-tonian}
\def\varb{\bih\ \var}

\def\ger{hierarch}

\def\omnman{$\omega N$ manifold}

\def\dncoo{Darboux-Nijenhuis coordinates}

\def\St{St\"ackel}

\newcommand{\finon}[2]{{\>#1=#2,\ldots,n}}
\newcommand{\fino}[3]{{\>#1=#2,\ldots,#3}}
\newcommand{\Lie}[1]{{\text{Lie}_{{#1}}}}
\begin{document}
\begin{flushright}
Ref. SISSA 106/2003/FM
\end{flushright}
\vspace{0.8truecm}
\begin{center}
{\Large\bf On Separation of Variables for }\\
{\Large\bf Homogeneous SL(r) Gaudin Systems}
\end{center}
\vspace{0.8truecm}
\begin{center}
{\large
Gregorio Falqui and Fabio Musso\\
SISSA, Via Beirut 2/4, I-34014 Trieste, Italy\\
falqui@sissa.it, musso@sissa.it}\\
\end{center}
\vspace{0.1truecm}
\begin{abstract}\noindent
By means of a recently introduced \bih\ structure for the homogeneous 
Gaudin models, we find a new set of Separation Coordinates for 
the $sl(r)$ case. 
\end{abstract}
\section{Introduction}
In this paper we will discuss the 
Gaudin system with $sl(r)$-valued spins
defined by the Hamiltonian
\begin{equation}
  \label{eq:1.1}
  H_G=\sum_{i<j}^n \text{Tr}(A_i\cdot A_j), \quad A_k\in sl(r)
\end{equation}
on the manifold $M=sl(r)^n$ equipped with
the standard product Lie--Poisson structure.
We will refer to it, with a slight abuse of notation, as the {\em  
homogeneous $sl(r)$ XXX Gaudin system}
(as in \cite{FaMu2}).

The ``conventional'' approach 
to the integrability of this quite well 
studied problem is based on the Lax representation and the 
$r$-matrix theory (see, e.g., \cite{Jurco,Sk89}).
Fixing $n$ distinct parameters $a_i,
\finon{i}{1}$ one introduces the matrix
\begin{equation}
  \label{eq:1.2}
  \CL(\la)=\sum_{i=1}^n \frac{A_i}{\la-a_i},
\end{equation} 
to be considered as an element of the Loop algebra $sl(r)((\la))$. 
Along the Hamiltonian flow defined by~\rref{eq:1.1},
the Lax matrix $\CL(\la)$ evolves according to a Lax equation
\[
\frac{d\,\CL(\la)}{dt} =[\CL(\la),M].
\]
Thanks to the existence of an $r$-matrix  
for the Lax matrix \rref{eq:1.2} the spectral invariants
\begin{equation}
  \label{eq:1.4}
  I^{(\al)}_i=\text{Res}_{\big\vert_{\la=a_i}}\text{Tr}(\CL(\la)^\al), 
\finon{i}{1},\fino{\al}{2}{r},
\end{equation}
are in involution. These integrals,  
together with the integrals of the motion associated with the 
invariance of the system under the
global 
$SL(r)$ action given by
\begin{equation}\label{eq:1.4a}
A_i\to GA_i G^{-1},
\end{equation}
to be referred to as {\em global gauge invariance},
provide a complete set of constants of the
motion for $H_G$.

The separability of the Hamilton--Jacobi equations associated with the Gaudin
Hamiltonian~\rref{eq:1.1} was first studied \cite{Sk89,Sk92},   for the 
low $r$ cases, as a kind of byproduct of the solution of
the Bethe Ansatz equations associated with the quantum Gaudin system. 
Separability was then proved for the general case in \cite{Gekhtman,Scott}.
and (implicitly) framed within the theory of 
Algebraically Complete Integrable Systems in \cite {AHH93,DiDu94}.    
In this scheme, it turns out that one can find a set of  
``algebro-geometrical'' Darboux coordinates $(\zeta_i,\la_i)$ as coordinates 
of a set of $d_n=r(r-1)(n-1)/2$
distinguished points on 
spectral curve
\begin{equation}\label{eq:1.genus}
\Gamma(\zeta,\la)=\text{Det}(\zeta-\CL(\la)),
\end{equation}
whose genus is $g=\dsl{\frac{(r-1)}{2}}\big((n-2)r+(r-2)\big)$.

In the recent paper~\cite{FaMu2} we have reconsidered the (homogeneous XXX) 
Gaudin model, and generalized to the case of an arbitrary 
Lie algebra $\alg$ an alternative set of
integrals of the motion for $H_G$ (see, e.g. \cite{Rag}), introduced in
the Hopf-algebraic approach to the integrability of the system.
The distinguished feature of such integrals, which in the case of $\alg=sl(2)$
are given by the very simple expressions
\begin{equation}
  \label{eq:1.5}
  K_l=\text{Tr}\Big(\sum_{i=1}^l A_l\Big)^2,
\end{equation}
is that they are independent of the (fake) parameters entering the definition
of the Lax matrix~\rref{eq:1.2}.

These integrals were also introduced, in a different context
\cite{KM}, as Hamiltonians of a notable class of
Hamiltonian flows on the  moduli space of $n+3$-sided polygons in
$\mathbb{R}^3$, and later generalized in ~\cite{FlMi01} to the $\mathbb{R}^d$
case\footnote{We thank J. Harnad for drawing our attention to these
  references.}. 
This moduli space turns out to be a suitable Marsden--Weinstein Hamiltonian
quotient of the Poisson manifold $su(2)^{n+3}$ associated with 
the corresponding Gaudin model. 
The Hamiltonian  flows associated with~\rref{eq:1.5}
were termed ``bending flows'' due to the 
following fact: if one draws, from a chosen vertex, the $n$ possible
diagonals of an $n+3$-sided polygon, the flow associated with the Hamiltonian  
$K_k$ geometrically represents the bending of one side of the polygon along
the $k$-th diagonal (the other side being kept fixed). 

The key point for the analysis performed in~\cite{FaMu2} was the introduction,
along with the standard Lie--Poisson structure $P$, 
of a particular second Poisson
structure, hereinafter called $R$. 
In the $n=3$ case, this structure is defined by its Hamiltonian vector
fields as follows:
\begin{equation}
  \label{eq:1.6}
  \left\{ \begin{array}{l} \dot{A_1}=[A_1,\ddd{F}{A_2}+\ddd{F}{A_3}]\\
 \dot{A_2}=[A_1,\ddd{F}{A_1}-\ddd{F}{A_2}]+[A_2,\ddd{F}{A_2}+\ddd{F}{A_3}]\\
 \dot{A_3}=[A_1,\ddd{F}{A_1}-\ddd{F}{A_3}]+[A_2,\ddd{F}{A_2}-\ddd{F}{A_3}]+
2[A_3,\ddd{F}{A_3}],\end{array}\right.
\end{equation}
where $\ddd{F}{A_i}$ are elements of $sl(r)$ to be properly defined in Section~\ref{sect:2}.

The Poisson pencil $R-\la P$ and  the integrals~\rref{eq:1.5}  
fulfill standard Lenard--Magri relations, namely one
can check that
\begin{equation}
  \label{eq:1.8}\begin{split}
  &P d\text{Tr}(A_i^2)=0, \finon{i}{1},\quad R d\text{Tr}(A_1^2)=0,\\
  &R d\text{Tr}\Big(\sum_{i=1}^a A_i^2\Big)
=P d K_a,\quad \finon{a}{2}.\end{split}
\end{equation}
For the general $sl(r)$ case one can show that it is possible to find a
sufficient number of polynomial functions
in involution that provide a set of integrals of the motion
$K_l^{(\al)}$ alternative to the set defined by the 
Lax matrix~\rref{eq:1.2}. They share with the integrals \rref{eq:1.5} the
property of being defined independently of the parameters entering the
Lax matrix~\rref{eq:1.2}.   

In the last Section of~\cite{FaMu2} we addressed the problem of separability of
such flows, in the framework of the so--called \bih\ approach to the SoV
problem (see, e.g., \cite{MT97,Bl98,MFP03,FP1}, 
and the references quoted therein). In particular we solved it for the
$sl(2)$ case by means of explicit computations, showing that the separation
coordinates associated with the pencil $R-\la P$ 
are rational functions of the natural coordinates $(h_i,e_i,f_i)$ in
$sl(2)^n$, and the separation relations are quadratic equations in these
coordinates. 

In this paper we will solve the corresponding problem for the $sl(r)$ case,
with $r$ arbitrary. This task will be accomplished by means of a careful
mixing of techniques of the theory of Lax equations with $r$--matrix structure,
and the theory of \bih\ systems such as those exposed in a series of papers by
Gel'fand and Zakharevich \cite{GZ} and Magri and collaborators 
\cite{FMT,MFP03}.
In particular, we will make extensive use of (refinements of) the results
presented in \cite{FP0,FP1} concerning the Separation of Variables of
systems with an arbitrary number of Lenard--Magri chains.
The key points for the analysis we are going to develop in this paper are:
\begin{enumerate}
\item It is possible to deform the Poisson tensor $R$ into $\tilde{R}$ in
  such a way that \\
a) $\tilde{R}$ is still compatible with $P$ and 
restricts to the (generic) symplectic leaf $\CS$  of $P$.\\
b) The integrals  $K_l^{(\al)}$ defined by the pencil $R-\la P$
are in involution also w.r.t. $\tilde{R}$,
although the recurrence relation they satisfy in relation with the new pencil
$\tilde{R}-\la P$ are more complicated than
the usual Lenard--Magri relations.  
\item It is possible to define $n-1$ Lax matrices $L_a$ linearly 
depending on a ``spectral parameter'' $\la$ such that any formal vector
field $X(\la)$ which is ``Hamiltonian w.r.t. the pencil $R-\la P$ induces a
Lax equation on each of the matrices $L_a$.
\end{enumerate}
Thanks to the first property it will be possible to endow the generic
symplectic leaf $\CS$ of $P$ with a special geometric 
structure, that is, a 
$(1,1)$ tensor with vanishing \Nij\  
torsion, whose ``spectral'' data will provide us with a set of separation
coordinates for the H--J equations associated with $H_G$.

Thanks to the second property, as well as other specific features of the
deformation $\tilde{R}$ of $R$, (to be fully discussed in the core of the
paper) we will be able to show that the separation relations are provided by
the spectral curves of the matrices $L_a$. The distinguished feature of such a
SoV scheme is that the separation coordinates are defined iteratively in
subsets of $d_r=r(r-1)$ coordinates, 
which are coordinates of a set of $r(r-1)/2$
points on a genus 
\[
g=(r-1)(r-2)/2
\]
curve, irrespectively of the number $n$ of
``sites'' of the Gaudin model.

A word of warning: the set of coordinates defined in this way on $\CS$ must be
completed by a set of $r(r-1)$ coordinates associated with the global $SL(r)$
invariance of the model, just like the set of integrals coming either from the
Lax matrix ~\rref{eq:1.2} or from the construction discussed in~\cite{FaMu2}
must be supplemented by the set of integrals associated with the global
gauge invariance of the model. However, since these integrals are
associated with a sort of ``cyclic'' coordinates, 
they will trivially enter the H-J
equations and the problem of separability. So in the core of the paper, we
will often ``forget'' about them.

The scheme of the paper is as follows: In Section \ref{sect:2} we will fix
some conventions and notations to be used throughout the paper, review
the results of~\cite{FaMu2} to be used in the sequel, and introduce the Lax
matrices $L_a$. In Section \ref{sect:3} we will briefly recall the main points
of the \bih\ scheme for SoV, and, 
in Section \ref{sect:4} we will discuss how to 
apply such a picture 
to the $sl(r)$
Gaudin models. Finally, in Section \ref{sect:5} we will 
give examples our constructions
in the $sl(2)$ and $sl(3)$ case. In the last section we briefly summarize the
content of the paper and add a few comments. In order to simplify the
presentation, we collected the proofs of some important but somewhat 
technical points in three Appendices.

\section{The bihamiltonian structures and the Lax matrices} \label{sect:2}
Let $\alg$ be the Lie algebra $sl(r)$. 
It is known that it (as well as any simple Lie algebra) 
can be identified with its dual, e.g.,
via the dual pairing given by the trace in the fundamental representation.
In this paper we will constantly use such an identification.
The Lie Poisson structure on $M=\alg^n$ is the one defined, in the natural
coordinates $\{A_1,\ldots,A_n\}$ by its Hamiltonian vector fields:
\begin{equation}
  \label{eq:2.1}
  \dot{A}_i=[A_i, \ddd{F}{A_i}],
\end{equation}
where, if $X=(X_i,\ldots,X_n)$ represents 
a tangent vector to
$M$, the elements $\ddd{F}{A_i}\in sl(r),\finon{i}{1}$ 
are those matrices  defined
by means of the expression of the Lie derivative of $F$ w.r.t. $X$ as
\begin{equation}
  \label{eq:2.2}
  \Lie{X}(F)=\sum_{l=1}^n \text{Tr}\big(X_i\cdot\ddd{F}{A_i}\big).
\end{equation}
We will hereinafter denote the Poisson tensor associated with the Lie-Poisson 
natural bracket by $P$. 
From,  e.g., \cite{MaMo84,RSTS}) 
we know that we can endow
$M$ with a multi-parameter family of Poisson structures which are compatible
with the natural one~\rref{eq:2.1}. In \cite{FaMu2} a further linear Poisson
structure, to be denoted by $R$, has been introduced. It can be described as
follows. 

We notice that relation~\rref{eq:2.1} can be written  as:
\begin{equation}
  \label{eq:2.3}
  \dot{A}_i=\sum_{j,k=1}^n p_{ijk}[A_k, \ddd{F}{A_j}],\quad\text{with }
  p_{ijk}=\delta_{ik}\delta_{ij}.
\end{equation}
The new Poisson tensor $R$ is analogously defined by the expression:
\begin{equation}
  \label{eq:2.3a}
   \dot{A}_i=\sum_{j,k=1}^n r_{ijk}[A_k, \ddd{F}{A_j}],
\end{equation}
with
``structure constants'' given by
\begin{equation}
  \label{eq:2.4}
  r_{ijk}=(k-1) \delta_{ij} \delta_{jk}-\theta_{(i-k)} \delta_{ij} +\theta_{(j-i)}
\delta_{ik}+\theta_{(i-j)}\delta_{jk}
\end{equation}
where $\delta$ is the usual Kronecker symbol and
$\theta_{(i)}$ equals $1$ if $i>0$, and vanishes for $i\le 0$. 
Explicitly, the Hamiltonian vector field associated by 
$R$ with a function $F$ is given by:
\begin{equation}
\dot{A}_i= [ A_i, (i-1)  \ddd{F}{A_i}+\sum_{k=i+1}^N \ddd{F}{A_k}]+\sum_{k=1}^{i-1} 
[ A_k, \ddd{F}{A_k}-\ddd{F}{A_i}]. \label{vf}
\end{equation}
The following facts can be proven~\cite{FaMu2}:
\begin{prop}\label{prop:c.i}
\begin{enumerate}
\item The pencil of bivectors $R-\la P$ is a \bih\ structure on $M$, that is,
  $R$ is a Poisson structure compatible with $P$. 
\item
the functions $H^{(1)}_{\al,1}=\frac1{\al+1}\text{Tr} (A_1^{\al+1})\>, 
\fino{\al}{1,}{r-1}$ 
are {\em common} Casimirs for
$R$ and $P$. The Lenard-Magri chains starting at
\[
H^{(a)}_{\al,1}= \frac{1}{\al+1}\text{Tr} (A_a^{\al+1}),\>\finon{a}{2}
\]
provide us with further $d=(n-1)r(r-1)/2$ functionally independent integrals
\begin{equation}\label{eq:2.6}
H^{(a)}_{\al,p}, \finon{a}{2},\fino{p}{2}{\al+1},\fino{\al}{1}{r-1}.
\end{equation}
\item Taking into account the integrals associated with the global $SL(r)$
  invariance of the model, that is the ring of functions generated by $
    F_\xi=\text{Tr}(\xi\cdot\sum_{i=1}^n A_i),$
those integrals insure complete Liouville integrability of the model.
\end{enumerate}
\end{prop}
\endpf
{\bf Remarks.} 
\begin{enumerate}\item
The Gaudin Hamiltonian ~\rref{eq:1.1} is expressed in terms of
the integrals~\rref{eq:2.6} as
\[
H_G=\sum_{a=2}^n H^a_{2,2}
=\sum_{a=2}^n \text{Tr}\big(A_a\cdot (\sum_{b=1}^{a-1}A_b)\big). 
\]
\item
A convenient choice of the integrals associated with the 
global $SL(r)$ invariance 
can be done as follows.  We pick the $r-1$ independent 
elements $F_{h_1},\ldots,F_{h_{r-1}}$ associated with, say, the standard 
Cartan subalgebra of $sl(r)$, and the Gel'fand-Cetlyn invariants, that is, 
the Casimirs of the nested subalgebras
\begin{equation}\label{eq:gece}
sl(2) \subset sl(3) \subset \dots \subset sl(r), 
\end{equation}
under the map 
$sl(r)^n\to sl(r)$ sending the $n$-tuple $\{A_1,\ldots,A_n\}$ into the total
sum, $A_{tot}=\sum_{i=1}^n A_i$.
\end{enumerate}
For the sequel of the paper the following construction is crucial.
Let us introduce  $n-1$ Lax matrices:
\begin{equation}\label{eq:s1}
L_a=(\lambda-(a-2))A_a + \sum_{k=1}^{a-1} A_k \qquad a=2,\dots,n,
\end{equation}
It holds:
\begin{prop}\label{prop:lax-ev}
Let F be a smooth function on $M$ and let us consider
the pencil of vector fields
\[
X_F^\la=P_{\lambda} d F:=(R-\la P) dF
\]
(we say that $X_F^\la$ is Hamiltonian w.r.t. the
pencil $P_\la$). Then, along $X_F^\la$, the  matrices $L_i$ of eq. \rref{eq:s1}
evolve according to a Lax equation,  
\begin{equation}\label{eq:s8}
\Lie{X_F^\la}(L_a)=[L_a(\lambda),M_a(\lambda)]
\end{equation}
with
\begin{displaymath}
M_a(\lambda)=(a-1-\lambda) \frac{\partial F}{\partial A_{a}}+
\sum_{b=a+1}^n  \frac{\partial F}{\partial A_{b}}
\end{displaymath}
\end{prop}
{\bf Proof:} Let us denote $\alpha_i=\frac{\partial F}{\partial
  A_{i}}, \finon{i}{1}$.
The vector field $X_F^\la$ is explicitly given by:
\begin{eqnarray*}
&&\Lie{X_F^\la}(A_i)=(P^{\lambda} d F)_i= \sum_{j,k} (r_{ijk}-\lambda p_{ijk})
\left[ A_k,  {\alpha_{k}} \right]=\\
&& = \sum_{j,k} \left( (k-\lambda-1)\delta_{ij} \delta_{jk} -
\theta(i-k) \delta_{ij} + \theta(j-i) \delta_{ik} + \theta(i-j) \delta_{jk}
\right) \left[ A_k,  {\alpha_{k}} \right]=\\
&& = \sum_{k=1}^{i-1}  \left[ A_k,  {\alpha_{k}}
-  {\alpha_{i}} \right]+ \left[ A_i, (i-\lambda-1) 
{\alpha_{i}}+ \sum_{k=i+1}^{N}  {\alpha_{k}}  \right]
\end{eqnarray*}
Substituting in $L_a(\lambda)$ we get:
\begin{eqnarray*}
\Lie{X_F^\la}(L_a)&=&(\lambda-a+2) \left( \sum_{k=1}^{a-1}  \left[ A_k,  {\alpha_{k}}-  {\alpha_{a}} \right]+ \left[ A_{a}, (a-1-\lambda) 
{\alpha_{a}}+ \sum_{k=a+1}^{N}  {\alpha_{k}}  \right] \right)+\\
&+& \sum_{j=1}^{a-1} \left( \sum_{k=1}^{a-2}  \left[ A_k,  {\alpha_{k}}
-  {\alpha_{j}} \right]+ \left[ A_j, (j-\lambda-1) 
{\alpha_{j}}+ \sum_{k=j+1}^{N}  {\alpha_{k}}  \right] \right)=\\
&=&   \left[\left((\lambda-a+2) A_{a}+ \sum_{j=1}^{a-1} A_j\right) , 
\left((a-1-\lambda) 
{\alpha_{a}}+ \sum_{k=a+1}^{N}  {\alpha_{k}} \right) \right]
\end{eqnarray*}
\endpf
We can interpret this result by saying that we can associate with the
homogeneous $n$-particle Gaudin system a set of $n-1$ matrices depending on a
parameter $\la$, satisfying a Lax equation along the ``formal'' (i.e.,
depending on the parameter $\la$) flows of
vector fields that are Hamiltonian with respect to the pencil
$P_\la$. 

\begin{prop}\label{prop:cotp}
The coefficients $\Ka{a}_\al(\la)$ of the expansion in powers of $\mu$ of the
characteristic polynomial 
\[
\text{det}(\mu-L_a(\la))=\mu^r+\sum_{\al=1}^{r-1} \Ka{a}_\al(\la)\mu^{r-\al-1}
\] 
of every Lax matrix $L_a(\la)$
are polynomial Casimirs of the pencil $P_\la=R-\la P$. Moreover, 
along any vector field $X$ associated with any of the non--trivial coefficient 
of such polynomial Casimir,
all matrices $L_a(\la)$ evolve according to Lax equations
\[
\Lie{X}( L_a(\la))=[L_a(\la),M_a(X)]
\]
for suitable matrices $M_a(X)$.
\end{prop}
{\bf Proof.} These assertion follow from the general theory of \bih\ pencils
on loop algebras (see, e.g.,~\cite{RSTS} 
and \cite{PV}). We sketch the proof for completeness,
considering the equivalent set of spectral invariants 
$\Ha{\al}_m=1/(\al+1) \text{tr}(L_m)^{\al+1}$.
To prove the first statement, we must show that,
for any one--form $v$, that we can
assume to be exact, $v=dF$ we have
\[
\langle v, P_\la d  \Ha{l}_m\rangle =0.
\]
Now, switching the action of the Poisson pencil on 
$v=dF$ the LHS of this equation reads
\begin{equation}
\begin{split}
L_{X_F^\la}( \Ha{l}_a)&=L_{X_F^\la}(1/(\al+1) \text{Tr}(L_a(\la))^{\al+1})\\&
=\frac{1}{\al+1}\sum_{p=0}^{\al-1}
\text{Tr}\left(L_a(\la)^p\cdot L_{X_F^\la}(L_a(\la))\cdot
  L_a(\la)^{\al-p}\right)\\
&=\text{Tr}\left(L_a^\al\cdot L_{X_F^\la}(L_a)\right)=
\text{tr}\left(L_a^\al(\la)\cdot[L_a(\lambda),M_a^F(\lambda)]\right)=0
\end{split}
\end{equation}
This proves the first assertion of the proposition, and, in particular, 
shows that all the vector fields $X^{(a)}_{\al,p}$ associated
(say, via $P$) with the coefficients of the expansion 
\begin{equation}\label{Hm}
\Ka{a}_\al=\sum_{p\ge 0}
\Ka{a}_{\al,p}\la^p 
\end{equation}
are indeed \bih\  vector fields.  

To prove the second statement we notice, using a very simple trick well 
known to experts in the \bih\ theory of integrable system, 
that $X^{(a)}_{\al,p}=P\Ka{a}_{\al,p-1}$ 
can be written as a Hamiltonian vector field
w.r.t. the pencil, considering the ``truncated'' 
polynomial $\left( \la^{-p}
  \Ka{a}_\al(\la)\right)_+$, 
where $(\cdot)_+$ denotes the nonnegative part of the expansion in $\la$. So
we see that the \bih\ vector fields of the \ger y are as well Hamiltonian
vector fields w.r.t. the \bih\ pencil $P_\la$.
The assertion then follows from Proposition \ref{prop:lax-ev}.
\endpf
Let us now focus our attention on a single Lax matrix, say $L_{\bar a}$;
calling, for simplicity, $B_{\bar a}=\sum_{b=1}^{\bar{a} -1} A_b$, we have that
the matrix 
\[
L_{\bar a}=(\la-\bar{a}+2)A_{\bar a}+B_{\bar a}
\]
is a Lax matrix with spectral parameter that evolve according to Lax equations 
along the vector fields of the hierarchy. Clearly, the Poisson brackets
induced on $M^{(2)}=sl(r)\times sl(r)$ by the map $M\to M^{(2)}$ defined by
$\{A_{\bar a}, B_{\bar a} \}$ are nothing but the Lie Poisson brackets on
$M^{(2)}$. So, applying the formalism of \cite{Sk89,AHH93,DiDu94},i.e., 
according to the Sklyanin ``magic recipe'' \cite{Sk95}, 
we can get, {\em for every fixed $\bar a$} a set of canonical 
coordinates $\{\xi^{\bar a}_\al,
\la^{\bar a}_\be\}$. Actually, we shall do this in Section \ref{sect:4}.

The point is that, to get a set of canonical coordinates
for the whole systems, we have to compare the different sets of coordinates
coming from the different Lax matrices $L_a, \finon{a}{2}$ (and those coming 
from the global gauge invariance of the model).

To solve this problem, we shall
make use of the \bih\ structure of the problem, and, namely, 
frame the Gaudin systems within the so--called \bih\ scheme for SoV.  
For the case of $sl(2)$, we
were able to solve the problem by means of straightforward computations. For
the general case, we have to use some slightly more sophisticated ideas and
techniques of the \bih\ theory, to be discussed in the next Section. 

\section{Bihamiltonian geometry and Separation of Variables}\label{sect:3}
As we already remarked in the Introduction,  a theory 
of Separation of variables based on the notions of \bih\
geometry has been quite recently introduced in the literature. 
The basic property of such a theoretical scheme  
which will enable us to solve the SoV problem of this paper
can very simply stated as follows:
\begin{prop}
Let $(M, P_1-\la P_0)$ be a \varb\ and suppose that there exist functions
$f,g,\lambda_f,\lambda_g$, with $\lambda_f\neq\lambda_g$, (eventually, $\la_f$
and/or $\la_g$ might be constant) satisfying 
\begin{equation}
  \label{eq:3.n1}
  P_1 df=\la_f P_0 df,\quad P_1 dg=\la_g P_0 dg. 
\end{equation}
Then $\{f,g\}_{0}=\{f,g\}_1=0$.
\end{prop}
{\bf Proof}. 
The assertion easily follows from the equations
\[
\begin{split}
&\{f,g\}_1=\langle df, P_1 dg \rangle=\la_g\{f,g\}_0\\
&\{g,f\}_1=\langle dg, P_1 df \rangle=-\la_f\{f,g\}_0,
\end{split}
\]
\endpf
In words, calling (with a slight abuse of language) a function $f$ satisfying
\rref{eq:3.n1} an ``{\em eigenvector}'' of the pair $P_1,P_0$ 
relative to the ``{\em eigenvalue}'' $\la_f$ as
in~\cite{HaHu02}, this proposition simply says that  eigenvectors 
belonging to different eigenspaces mutually commute. 

If the Poisson tensors $P_1$ and $P_0$ do not share the same image and kernel,
then a complete set of eigenvectors cannot be found. This is a typical
instance in the Gel'fand-Zakharevich theory of \bih\ integrable systems, (and
happens for the Poisson tensors $R$ and $P$ that we have considered so far).  

In general, the \bih\ theory of SoV suggests to consider a suitable
deformation $\wid{P}_1$ of $P_1$, such that 
it restricts to the (generic) symplectic leaves of $P$ 
and it is still compatible with $P$. 
Upon restriction, the generic symplectic 
leaf $\CS$ of $P$ will be endowed with a regular \bih\
structure (that is, a \bih\ structure in which one element of the pencil is
invertible). So, in the terminology of~\cite{FP1}  
the generic symplectic leaves of $P$ are 
{\em \omnman s},  that is are symplectic manifolds (with symplectic form
naturally induced by $P$), endowed 
with a compatible \Nij\ (or
hereditary) tensor $N$.  In terms of the Poisson structures,  
the \Nij\ tensor on the symplectic leaf $\CS$ is defined by  
\[
N=\wid{P}_1{}\vert_\CS\cdot P_0^{-1}\vert_{\CS}. 
\]

To concoct out of $P_1$ the suitable deformation $\wid{P}_1$, 
one can adopt the
following strategy:\\
First one fixes a complete set $C_1, \ldots, C_k$ of Casimirs of $P_0$,
considers the first vector fields of the Lenard chains associated with
$C_a$, i.e.,  
\[
X_a=P_1 dC_a, a=1,\ldots,k,
\]
and a a distribution $\CZ$, transversal to the symplectic leaves of $P_0$.
For any basis $W_1, \ldots, W_k$ in $\CZ$, the matrix
\begin{equation}
  \label{eq:n2}
  [G_0]_{a,b}=\Lie{W_b}(C_a)
\end{equation} is nonsingular (say on an open set $U\subset M$).
So, the tensor defined by
\begin{equation}
  \label{eq:3n}
\wid{P}_1=P_1-\sum_{a,b} X_a\wedge [G^{-1}_0]_{a,b} W_b
\end{equation}
is well defined and restricts to the generic symplectic leaf $\CS$ of $P$,
since, by construction, $\wid{P}_1 dC_j=0, \fino{j}{1}{k}$.
Notice that, if we define a new basis in $\CZ$ by
\begin{equation}
  \label{eq:4n}
  Z_a=\sum_b[G_0^{-1}]^a_b W_b, \text{ so that }\Lie{Z_a}(C_b)=\delta_{a,b},
\end{equation}
the expression of the deformed tensor $\wid{P}_1$ simplifies 
to 
\begin{equation}\label{eq:3.op}
\wid{P}_1=P_1-\sum_{a} X_a\wedge Z_a.
\end{equation}
We will call a basis of $\CZ$ satisfying
~\rref{eq:4n} a {\em normalized basis for the transversal distribution}. 

The proof of the following Proposition can be found in~\cite{FP0,DM}
\begin{prop}\label{prop:transdist} 
Let $(M,P_1-\la P_0)$ be a $2n+k$ dimensional 
\varb\  with $\text{corank}(P_0)=k$, 
and suppose
that there exists an integrable 
distribution $\CZ\subset TM$ of dimension $k$, s.t.:
\begin{enumerate}
\item $\CZ$  intersect
transversally the symplectic foliation of $P_0$.
\item\label{it2} the space of
functions invariant under $\CZ$ is a Poisson subalgebra for the whole pencil
$(M,P_1-\la P_0)$.
\end{enumerate}
Then, if $\wid{P}_1$ is the deformation of $P_1$ defined by~\rref{eq:3n},
$\wid{P}_1-\la P_0$ is still a Poisson pencil on $\CM$, and its
restriction endows the generic symplectic leaves of $P_0$ 
with the structure of a \omnman.
Furthermore, if  $Z_1,\ldots,Z_k$ are a set of generators of $\CZ$, 
normalized  w.r.t. a given complete set $C_1,\ldots,C_k$ of Casimir 
functions of $P_0$,condition \ref{it2} above translates into the equations:
  \begin{equation}
    \label{eq:vais}
    \Lie{Z_a}P_0=0,\quad \Lie{Z_a}P_1=\sum_{b=1}^k [Z_a,X_b]\wedge Z_b,\quad\text{where } 
    X_a=P_1 dC_a, \quad a=1,\ldots,k.
  \end{equation}
\end{prop}
\endpf

\begin{defi}\label{def:affine}
We say that a \bih\ manifold $(M,P_1-\la P_0)$, endowed with a transversal 
distribution $\CZ$
satisfying the assumptions of Proposition \ref{prop:transdist} admits an affine
structure if it is possible to choose a complete set of Casimir of $P_0$, and
a corresponding basis of {\em normalized flat generators}  
$Z_b,\fino{b}{1}{k}$ 
in $\CZ$ such that, for every Casimir of the
Poisson pencil $H^a(\la)$ and every $b,c=1,\ldots k$ 
one has, in addition to Equation
\rref{eq:vais}
\begin{equation}
  \label{eq:3.x}
  \Lie{Z_b}\Lie{Z_c}(H^a(\la))=0.
\end{equation}
\end{defi}
The notion of affine structure for a \varb\ 
was studied in~\cite{FP1} in connection with the problem of the \St\
separability of a \bih\ system. For the purposes of the present paper, we
remark that an affine Poisson pencil satisfies special properties, to 
be illustrated in the following.  

Let $(M, P_1-\la P_0)$ be a corank $k$ affine \bih\  manifold, and let $Z_a,
\fino{a}{1}{k}$ be a set of normalized flat generators for the transversal distribution
$\CZ$. Let us consider the polynomial Casimirs
\[
\Ha{a}(\la)=\la^{n_a}H^a_0+\cdots+H^a_{n_a},
\]
and their deformations along the flat generators, that is, the $k^2$
polynomials 
\begin{equation}
  \label{eq:3.2.D}
D^a_b(\la)=\Lie{Z_b}\Ha{a}(\la)=
\la^{n_a}\delta^a_b-D^a_{b,\, 1}\la^{n_a-1}-\cdots-D^a_{b,\,n_a}.
\end{equation}
The polynomials $D^a_b(\la)$ are invariant along $\CZ$, so that
they can be considered  as functions on
the generic symplectic leaves of $P_0$. They are the building blocks of the
\bih\ set-up for SoV for GZ systems.
Indeed it holds:
\begin{prop}\label{prop:delta1}
Let $\Delta(\la)$ be the determinant of the matrix $D^a_b$ of \rref{eq:3.2.D}.
then
\begin{enumerate} 
\item  The roots $\la_i$ of $\Delta(\la)$ satisfy
\[
\wid{P}_1 d\la_i=\la_i\, P_0\,d \la_i,
\]
\item
Let $\wid{D}(\la)$ denote the classical adjoint matrix of ${D}(\la)$, and let
$[\wid{D}(\la)]_{a,c}$ be non identically vanishing. 
Then any  ratio
$\rho(\la):=[\wid{D}(\la)]_{a,b}/[\wid{D}(\la)]_{a,c}$ 
of elements belonging to the $a$-th row of $\wid{D}(\la)$,   
evaluated at the roots $\la_i$ of $\Delta(\la)$ satisfy the equation
\begin{equation}
  \label{eq:3.5n}
  \wid{P}_1 d\rho(\la_i)=\la_i P_0d\rho(\la_i).
\end{equation}
\end{enumerate}
\end{prop}
The proof of this Proposition is contained in Appendix A.
\endpf
\subsection{Separation of Variables for the $\mathbf{sl(r)}$ 
Gaudin Systems}\label{sect:4}
In this subsection we will specialize the results 
of Section \ref{sect:3} and show how
the \bih\ structure $P_\la=R-\la P$ associated in Section \ref{sect:2} 
with the parameter independent
integrals of the Gaudin model provides a set of separation coordinates and
relations for the H-J equations associated with $H_G$.

The first step is to show that $P_\la$ induces a \omnman\ structure on the
generic symplectic leaf $\CS$ of $P$, that is, that the tensor $R$ can be
suitably deformed. 
We consider in $M=sl(r)^n$ 
the $n(r-1)$ vector fields 
\begin{equation}
  \label{eq:b4}
  W_i^\al:=\ddd{}{[A_i]_{r,\al}},\quad i=1,\dots,n,\quad \al=1,\dots,r-1,
\end{equation}
that is, the vector fields defined by their action on the $n$-tuple of
matrices $(A_1,\ldots, A_n)$ by
\begin{equation}\label{eq:nuova1}
\Lie{ W_i^\al}(A_1,\ldots, A_n)=(0,0,\cdots,\underbrace{e_{\al,r}}_{\text{i-th
    place}},\cdots,0),
\end{equation}
where $e_{\al,r}$ is the elementary matrix $(e_{\al,r})_{ij}=\delta_{i,\al} \delta_{j,r}$.
\begin{prop}\label{prop:b2}
The distribution $\CZ$ spanned by the vector fields $ W_i^\al$ satisfies
the hypotheses of Proposition \ref{prop:transdist}.
\end{prop}
The proof of this Proposition is contained in Appendix B. 
\endpf

To construct a set of flat generators $Z^\al_i$ for $\CZ$, we 
can argue as follows. In the case of a single copy of $sl(r)$, we normalize
the $W^\al$ with respect to the coefficients $C_1,\ldots,C_{r-1}$ of the 
characteristic
polynomial of $A$. 
The normalization for the $n$ site case
is done site by site. 
Since the determinant of a matrix is a linear function of each of its entry,
it is not difficult to realize that such normalized
generators $Z_i^\al$ provide the GZ manifold $M, R-\la P$ with the structure of
an affine GZ manifold, according to Definition~\ref{def:affine}. 

Let us now consider the Lax matrices
\[
L_a=(\la-a+2)A_a+\sum_{b=1}^{a-1} A_b, \> \finon{a}{2}, \text{ as well as }
L_1=A_1. 
\]
Define  $M_i(\lambda,\xi)=\xi\mathbf{I}-L_i(\lambda)$,
and denote their classical adjoint with $\wid{M}_i(\la,\xi)$.
The determinants of the matrices $M_i$ define, thanks to 
Proposition~\ref{prop:cotp}, 
polynomial Casimirs $\Ka{i}_\al(\la)$ for $R-\la P$, via:
\begin{equation}
  \label{eq:4.1.1}
  \text{Det}(M_i(\la,\xi))=\xi^r+\sum_{\al=1}^{r-1}\Ka{i}_\al(\la) 
\xi^{r-\al-1}.
\end{equation}
In particular, $\Ka{1}_\al$ are the common Casimirs of $P$ and $R$, while
$\Ka{a}(\la)$ are, for $\finon{a}{2}$, 
the non trivial polynomial Casimirs of the
pencil $R-\la P$.
 
Let us consider the $n^2$ matrices $D_{ij}$ defined by
\begin{equation}\label{eq:nuova2}
\big(D_{ij}\big)^\al_\be=\Lie{Z^\al_j}(\Ka{i}_\be(\la)).
\end{equation}
The proof of the following Proposition, which is based on a few elementary 
properties of the matrices $\big(D_{ij}\big)$  is contained in Appendix C.
\begin{prop}\label{prop:penult}
The determinant $\CD_a(\la)$ of the matrices $\big(D_{aa}\big), a=2,\ldots,
n$, factors as 
\begin{equation}
  \label{eq:nuova3}
  \CD_a(\la)=(\la-a+2)^{r-1}\Delta_a(\la),
\end{equation}
where $\Delta_a(\la)$ is a monic polynomial of degree $r(r-1)/2$.
Let  $\la^s_a, a=2,\ldots,n, s=1,\ldots,
r(r-1)/2$, be the roots of $\Delta_a(\la)$, and let us consider 
a row (say, the first) 
$\delta_\al(\la)=\big(\wid{D_{aa}}(\la)\big)_{1,\al}$ of the adjoint matrix
$\wid{D_{aa}}(\la)$ of $D_{aa}(\la)$. Finally, let 
$\xi^s_a$ be the functions obtained by evaluating in 
$\la=\la^s_a$ the ratios $\delta_{r-2}(\la)/\delta_{r-1}(\la)$.
Then, 
these $(n-1)r(r-1)/2$ pairs of functions $\{\xi^s_a,\la^s_a\}$ satisfy \\
1)
the Jacobi separation relations 
\[
\text{Det}(M_a(\xi^s_a,\la^s_a))=0.
\]
2) the differential relations 
\[
  \wid{R}d\la^s_a=\la^s_a Pd\la^s_a,\quad   
\wid{R}d\xi^s_a=\la_a^s Pd\xi^s_a.
\]
In particular, their brackets, (say, with respect to the Lie Poisson structure
$P$), are of the separate form:
\begin{equation}
  \label{eq:4.1.pb}
  \{\la_b^s,\xi_a^t\}_{P}=\delta^{st}\delta_{ab} 
\varphi^{s}_{a}(\xi_a^s,\la_a^s),
\end{equation}
where $\varphi^{s}_{a}(\xi_a^s,\la_a^s)$ are
functions of the two variables $(\xi_a^s,\la_a^s)$.
\end{prop}
\endpf
The meaning of this Proposition can be rephrased as follows. For every integer
$\finon{a}{2}$ we can construct a Lax matrix, whose characteristic polynomial
gives us a family of Casimirs of the \bih\ pencil $R-\la P$ defined on the
manifold $M=sl(r)^n$. Separated coordinates are constructed, 
according to the \bih\ scheme, by deforming such
Casimirs along normalized generators of a suitable distribution $\CZ$ defined
in $M$. In particular, for each $\finon{a}{2}$, we can algebraically 
construct a ``cluster'' of $(r(r-1))$ variables
$\{\la_a^s,\xi_a^s\}_{s=1,\ldots r(r-1)/2}$ that are, in the terminology of
\cite{HaHu02}, algebro-geometrical \Nij\ coordinates, that is, satisfy
properties 1) and 2) of Proposition \ref{prop:penult}.

To finish our job we have to:\\
i) Discuss about the coordinates associated with the global gauge invariance
of the Gaudin Systems\\
ii) Explicitly construct, out of the coordinates found so far, a set of
{\em canonical} separated coordinates (that is, a set of \dncoo).\\
Point i) can be solved as follows. One notices that any function $\varphi$ 
depending
only on the ``global'' matrix variable $A_{T}=\sum_{i=1}^n A_i$, which is
invariant along the distribution $\CZ$ satisfies the differential relation
\begin{equation}
  \label{eq:4.2.gl}
   \wid{R}d\varphi=(n-1)P d\varphi.
\end{equation}
In particular, this family includes the mutually commuting Hamiltonians of
Gel'fand-Cetlyn type discussed in the Remark after Proposition \ref{prop:c.i}.
The property~\rref{eq:4.2.gl} follows from the fact that they 
trivially satisfy the relation $Rd\varphi=(n-1)P d\varphi$
w.r.t. the undeformed pencil, and from the property that $\varphi$ commutes
with all the Hamiltonians of the hierarchy. 
Inside this ring of functions one can find a set of $r(r-1)/2$
canonical coordinates that complement the Gel'fand-Cetlyn Hamiltonians. 
Thanks to~\rref{eq:4.2.gl} they will have vanishing
Poisson brackets with the \Nij\ coordinates of Proposition \ref{prop:penult}.

The solution to point ii) above can be simply done by means of a direct
computation of the Poisson brackets between $\xi_a^s$ and $\la^s_a$. 
In particular, this computation will implicitly prove that these quantities
are functionally independent.
\begin{prop}\label{lem:fm}
The Poisson brackets, w.r.t. the Lie Poisson pencil $P$ of
the coordinates $\lambda_a^s,\xi_a^s$ defined above are given by 
\begin{equation}  
\{\lambda^s_a,\xi^s_a \} =(\lambda^s_a-a+2)(\lambda^s_a-a+1) \label{poi}
\end{equation}  
\end{prop}
{\bf Proof.}
The proof of this proposition follows verbatim that of Theorem
$1.3$ in \cite{AHH93}, to which we refer for full details. 
Indeed, the coordinates $\lambda_a^s,\xi_a^s$ can be seen as common zeroes
of the first row of the matrix $\widetilde{M}_a(\lambda, \xi)$. So we can
apply all the considerations of \cite{AHH93}, the only difference being that
the Poisson brackets  of the entries of the matrix $M_a(\lambda,\xi)$ are given by:
\begin{eqnarray}
&& \{ M_a^{ij}(\lambda,\xi),  M_a^{kl}(\sigma,\eta) \}= \nonumber \\
&& ={\rm{tr}} \left[ \left( (\lambda-a+2)(\sigma-a+2) A_a+ 
\sum_{r=1}^{a-1}A_r \right) 
\left(e_{kj}\delta_{li}-e_{li} \delta_{kj} \right) \right]= \nonumber\\
&& =\frac{1}{\lambda-\sigma} \left[ (\lambda-a+1)(\sigma-a+2)(
  M_a^{jk}(\lambda,\xi) 
\delta_{il}
-  M_a^{il}(\lambda,\xi) \delta_{jk})+ \right.  \nonumber \\
&&+ \left. (\lambda-a+2)(\sigma-a+1)(  M_a^{il}(\sigma,\eta) \delta_{jk}- M_a^{jk}(\sigma,\eta) \delta_{il}) \right].
\end{eqnarray} 
The presence of the factors $(\lambda-a+2)(\sigma-a+1)$ and
$(\lambda-a+2)(\sigma-a+1)$ is responsible for the factor
$(\lambda^s_a-a+2)(\lambda^s_a-a+1)$ in Eq. ~\rref{poi}.
\endpf
\section{Examples}\label{sect:5} 
In this Section we will specialize the constructions 
presented in the paper for the cases of $sl(2)$ and $sl(3)$.
\subsection{The $\mathbf{sl(2)}$ case}\label{subs5.1}
Here we briefly reframe the explicit computations of the last Section of  
\cite{FaMu2} within the formalism exposed in this paper. 
We consider the manifold $M={sl(2)}^{n}$, endowed with 
the Poisson pencil $R-\la P$ of Section \ref{sect:2}. It
is explicitly parametrized in terms of the $n$ matrices
\begin{equation}
A_i=\left[\begin{array}{cc} h_i& e_i\\ f_i&-h_i\end{array}\right].
\end{equation}
The generic symplectic leaf $S$ of $P$ is a  
$2n$ dimensional symplectic manifold, defined by the equations
\begin{displaymath}
C_i=\frac12\text{Tr}A_i^2=h_i^2+e_i f_i, \quad  i=1,\ldots, n,
\end{displaymath}
and can be (generically) endowed with the $2n$ coordinates $(h_i,
f_i), \> {i=1,\ldots, n}$.

A set of normalized transverse vector fields are given in this case by
\begin{equation}\label{eq:s3}
Z_i=\frac{1}{f_i}\ddd{}{e_i},
\end{equation}
The matrices $L_a$ are explicitly given by
\begin{equation}
L_a(\la)=\left(\begin{array}{cc}
(\la-a+2)h_a+\sum_{b=1}^{a-1}h_b&(\la-a+2)e_a+\sum_{b=1}^{a-1}e_b\\
(\la-a+2)f_a+\sum_{b=1}^{a-1}f_b&-((\la-a+2)h_a+\sum_{b=1}^{a-1}h_b)
\end{array}
\right).
\end{equation}
As canonical coordinates associated with the global $SL(2)$ invariance one can
choose the two functions
\[
\la_1= \sum_{i=1}^n f_i,\quad \phi_1=\frac{\sum_{i=1}^n h_i}{\sum_{i=1}^n f_i}.
\]
The non trivial separation coordinates are gotten simply considering
the zeroes $z_a$ of the elements $[L_a]_{2,1}$, and the values $\mu_a$ 
on these zeroes of the elements $[L_a]_{2,2}$, normalized as in the previous
Section. One sees that
\[
z_a= -\frac{ \sum_{k=1}^{a-1} f_k}{f_a}+(a-2), \quad a=2,\dots, n
\]
Shifting these values by the unessential term $a-2$, we find that the
separation coordinates are given, for $ a=2,\ldots, n$, by
\begin{equation}\label{eq:sl2sepcoo}
\lambda_a= -\frac{\sum_{k=1}^{a-1} f_k}{f_a}, \quad
\mu_a =- \frac{\lambda_a h_a+ \sum_{k=1}^{a-1} h_k}{
\la_a(\la_a-1)}
\end{equation}
They fulfill the separation relations
\begin{equation}\label{eq:sl2seprel}
\mu_a^2=\frac{1}{2(\la_a(\la_a-1))^2}
\big(C_a^2\la_a^2+\text{Tr}(A_a(\sum_{b=1}^{a-1}A_b))
\la_a+\text{Tr}((\sum_{b=1}^{a-1}A_b)^2)\big).
\end{equation}
In other words, the separation coordinates are coordinates of suitable points
on the rational curves \rref{eq:sl2seprel}. The corresponding Hamilton-Jacobi
equations can be explicitly solved by means of algebraic functions. 

\subsection{The $\mathbf{sl(3)}$ case}\label{subs5.2}
We consider the Poisson manifold $M=sl(3)^n$, endowed
with the Poisson pencil $R-\la P$ and parametrized by the $n$ matrices
\begin{equation}
  \label{eq:5.1}
  A_i=\left(\begin{array}{ccc} h_{1,i}& e_{1,i}&e_{3,i} \\
                               f_{1,i}& h_{2,i}-h_{1,i}& e_{2,i}\\
                               f_{3,i}&f_{2,i}&-h_{2,i}\end{array}
\right)\>\finon{i}{1}.
\end{equation}
On this manifold the Poisson tensor $P$ has $2n$ Casimirs:
\begin{equation}
C_i^2=\frac12 \text{Tr}\big((A_i^2)\big),\quad C_i^3=\frac13
\text{Tr}\big((A_i)^3\big) \quad i=1,\ldots,n. \label{CC}
\end{equation}
The characteristic polynomials of the 
Lax matrices 
$ L_a=(\la-a+2)A_a+\sum_{b=1}^{a-1} A_b$ are expressed as
\begin{equation}
  \label{eq:5.3}
  \Gamma^a(\mu,\la)=\mu^3-\mu\Ha{a}_2(\la)-\Ha{a}_3(\la).
\end{equation}
The transversal distribution $\CZ$ is generated by the $2n$ flat generators:
\begin{eqnarray*}
&&  Z^2_i=\frac{1}{d} \left[ (f_{3,i}(h_{1,i}-h_{2,i}) + f_{2,i} f_{1,i})\ddd{}{e_{2,i}}
+(f_{2,i}h_{1,i}-f_{3,i}e_{1,i})\ddd{}{e_{3,i}} \right] \\
&& Z^3_i=\frac{1}{d} \left[ f_{2,i} \ddd{}{e_{3,i}}-f_{3,i} \ddd{}{e_{2,i}} \right]\\
&& d=f_{2,i} f_{3,i} (2 h_{1,i}-h_{2,i}) + f_{2,i}^2 f_{1,i}-f_{3,i}^2 e_{1,i}
\end{eqnarray*}
The symplectic leaves of $P$ are generically parametrized by matrices $A_i$ of
the form:
\[
\left(\begin{array}{ccc} h_{1,i}& e_{1,i}&\Phi_{3,i} \\
                               f_{1,i}& h_{2,i}-h_{1,i}& \Phi_{2,i}\\
                               f_{3,i}&f_{2,i}&-h_{2,i}\end{array}\right)
\]
where $\Phi_{2,i}$ and $\Phi_{3,i}$
are suitable functions of the coordinates
$h_{1,i}, h_{2,i},f_{1,i},f_{2,i},f_{3,i}, e_{1,i}$, parametrically depending
on the Casimirs (\ref{CC}).

The  coordinates $\{\la_a^s,\xi_a^s\}$ can quite explicitly 
be found by means of the following steps:

We consider the matrix $M_a(\la,\xi)=\xi-L_a(\la)$ and its adjoint 
$\widetilde{M}_a(\la,\xi)$.  We have to look for the common zeroes of the
elements $ \widetilde{M}_a(\la,\xi)_{3,1}$ 
and $\widetilde{M}_a(\la,\xi)_{3,2}$,
that is, for the common zeroes of 
\begin{equation}
  \label{eq:5.5}
  \text{Det}\left(\begin{array}{cc} -L_a(\la)_{2,1}& \xi-L_a(\la)_{2,2}\\
-L_a(\la)_{3,1}&-L_a(\la)_{3,2}\end{array}\right),\qquad 
\text{Det}\left(\begin{array}{cc} \xi-L_a(\la)_{1,1}& -L_a(\la)_{1,2}\\
-L_a(\la)_{3,1}&-L_a(\la)_{3,2}\end{array}\right).
\end{equation}
Taking into account the form of the vector fields $Z^\al_a$ and of the
characteristic polynomial \rref{eq:5.3}, we can identify the
system~\rref{eq:5.5} with
\begin{equation}
  \label{eq:5.s}
  \left\{ \begin{array}{c} \xi\Lie{Z^2_a}\Ha{a}_2+\Lie{Z^2_a}\Ha{a}_3=0\\
\xi\Lie{Z^3_a}\Ha{a}_2+\Lie{Z^3_a}\Ha{a}_3=0\end{array}\right.,
\end{equation}
where
\begin{equation}\label{eq:zzz}
\text{Det}(M_a(\xi,\la))=\xi^3-\Ha{a}_2\xi-\Ha{a}_3.
\end{equation}
As we have noticed in Section \ref{sect:4} , we can factor out 
$(\la-a+2)$ from each line of this system, and consider, in matrix form the 
equivalent system:
\begin{equation}
  \label{eq:5.m}
 \begin{array}{c} \langle\xi,1|\\ \hspace{1mm} \end{array}\left(\begin{array}{cc}
      G^a_{2,2}&G^a_{2,3}\\G^a_{3,2}&G^a_{3,2}\end{array}\right),
\quad\text{with} G^a_{\al,\beta}=\Lie{Z^\al_a}\Ha{a}_\beta/(\la-a+2).
\end{equation}
We notice that $G^a_{\al,\beta}$ are polynomials in $\la$ of degree $\al-1$,
so that the three zeroes $\la^1_a,\la^2_a,\la^3_a$ of the equation
\begin{equation}\label{eq:5.boh1}
\Delta_a(\la)=\text{Det}\left(\begin{array}{cc}
      G^a_{2,2}&G^a_{2,3}\\G^a_{3,2}&G^a_{3,2}\end{array}\right)=0
\end{equation}
are the compatibility
condition for the system \rref{eq:5.s}; the corresponding coordinates
$\xi^1_a,\xi^2_a,\xi^3_a$ are thus given by, e.g., 
\begin{equation}\label{eq:5.boh2}
\xi^s_a=-G^a_{3,3}/G^a_{2,3}\big\vert_{\la=\la_a^s}
\end{equation}
We remark that our procedure for finding separation coordinates exactly
matches the one introduced, in the framework of $r$-matrix theory, in
~\cite{DiDu94}.

Finally, defining $\zeta_a^s=\la^s_a-a+2, \finon{a}{n}$ 
and considering the pairs
\[
\{\zeta_a^s,\>
\rho_a^s\}=\xi_a^s/\zeta_{a}^s(\zeta_a^s-1), s=1,2,3,  \finon{a}{2},
\]
we see that the solution $W$ of the (stationary) 
Hamilton-Jacobi equations of the $sl(3)$ Gaudin 
model can be expressed as:
\begin{equation}
  \label{eq:5:W}
  W=\sum_{a=2}^n\left(\sum_{s=1}^3 \int^{P^s_a} 
\frac{\xi d\zeta}{\zeta(\zeta-1)}\right)+\sum_{\al=s}^3 H^s_T q^s_T,
\end{equation}
where $P^\al_a$ denotes the point $(\xi_a^\al, \zeta_a^\al)$ on the genus $g=1$
algebraic curve defined by equation~\rref{eq:zzz}, and
$H^s_T$ and $q^S_T$ denote, respectively, a suitable complete family of 
Gel'fand-Cetlyn
Hamiltonians associated with the global $SL(3)$ invariance of the model, and
their canonically conjugated variables.

Standard arguments show that the linearization of the flows associated with the
mutually commuting Hamiltonians we have considered in this paper, and hence
also the flow associated with the ``physical'' Hamiltonian of the $sl(3)$ 
Gaudin system  can be achieved by means of the Abel maps associated with the
differentials (the last four being of the third kind)
\begin{equation}
  \label{eq:5.abe}
 \frac{d\zeta}{\Gamma_\xi^a}, \> \frac{\zeta d\zeta}{(\zeta-1)\Gamma_\xi^a},
\>\frac{d\zeta}{(\zeta-1)\Gamma_\xi^a},\>
 \frac{
  d\zeta}{\zeta(\zeta-1)\Gamma_\xi^a},\>\frac{\xi
  d\zeta}{\zeta(\zeta-1)\Gamma_\xi^a},
\end{equation}
where $\Gamma_\xi^a=\dsl{\ddd{\text{Det}(M_a(\xi,\la))}{\xi}}$.
The case of $sl(r), \> r>3$ can be treated analogously.

\section{Conclusion and discussion}\label{sect:6}
In this paper we have reconsidered the SoV problem for the (homogeneous XXX)
Gaudin systems based on the Lie algebras $sl(r)$, from a particular 
standpoint. Namely, we considered the $n$-site system as a 
Gel'fand-Zakharevich system defined on the manifold $M=sl(r)^n$, with 
respect to the Poisson pencil $P_\la=R-\la P$, where $P$ is the usual Lie 
Poisson bracket on $M$, while $R$, given by  equations \rref{eq:2.3a} and
\rref{eq:2.4} is a further linear Poisson structure on $M$ that
was introduced in \cite{FaMu2}.

We showed that the system admits a set of $n-1$ ``Lax'' matrices, linear in
the spectral parameter $\la$ that evolve according to Lax equations along any
vector field that is Hamiltonian with respect to the Poisson pencil $P_\la$.

Thanks to this property, by using the \bih\ set-up for SoV, we managed to
define a set of separating coordinates, quite explicitly given in equations
\rref{eq:sl2sepcoo} for the $sl(2)$ case and by equations \rref{eq:5.boh1} and
\rref{eq:5.boh2} for the $sl(3)$ case. 

We notice that this set of coordinates provide an alternative set
of separation coordinates for the Hamilton--Jacobi equations associated with
the Gaudin Hamiltonian w.r.t. the coordinates that can be found by means of
the conventional approach based on the rational Lax matrix \rref{eq:1.2}, i,e,
\begin{equation}
  \label{eq:6.2}
  \CL(\la)=\sum_{i=1}^n \frac{A_i}{\la-a_i}.
\end{equation} 
In this respect, a few remarks are in order:

First we notice that the existence of different sets of separation coordinates
is to be ascribed to the {\em super-integrability} of the system. This is
particularly clear in the \bih\ setting, where the separation coordinates are
obtained, by means of the procedure outlined in Section~\ref{sect:3}, from the
polynomial Casimirs of a Gel'fand-Zakharevich Poisson pencil. 
The general problem of the
connections between super-integrability and ``multi-separability'' is, to the
best of our knowledge, still an open problem (see, \cite{Mil00} and the
references quoted therein). In particular, the classification problem for
these systems have been solved only for systems with a small number of degrees
of freedom. The homogeneous XXX Gaudin models are systems with
an arbitrarily high number of degrees of freedom where the connection between
super integrability and multi-separability happens, and their study
might shed light on the structural properties of this phenomena.

The second remark is the following. The SoV scheme based on the Lax
matrix~\ref{eq:6.2} leads to the definition of a divisor of 
degree 
\[
d_R=r(r-1)(n-1)/2
\]
on
the spectral curve $R(\la,\mu)=\text{Det}(\mu-\CL(\la))$ (see, e.g.,
\cite{Har00}). It is not difficult to ascertain that the genus of this
spectral curve is, for $\alg=sl(r)$ 
\[
g_\CL= \frac{(r-1)}{2}\big((n-2)r+(r-2)\big),
\]
that is, it grows linearly with the number $n$ of sites of the model.

As we have shown in Section \ref{sect:4} and exemplified in Section
\ref{sect:5} for $r=2,3$, the SoV scheme herewith outlined leads to 
consider the set of $n-1$ Lax matrices $L_a(\la)$ given by 
\begin{equation}
\label{eq:6.3}
L_a(\la)=(\la-a+2)A_a+B_a,\quad B_a=\sum_{b=1}^{a-1} A_b. 
\end{equation}
The separation coordinates parametrize sets of degree $r(r-1)/2$ divisors on
the spectral curves $\Gamma^a(\la,\mu)=\text{Det}(\mu-L_a(\la))$. We see that 
the genus of such curves is
\[
g_{\Gamma^a}=(r-1)(r-2)/2,
\]
that is depends only on the rank $r$ of the algebra, and not on the number $n$
of sites, showing that the equations of motions can be explicitly solved by
means of $\theta$ functions of genus $g_{\Gamma^a}$ for all $n$'s.

This also implies the existence of  canonical transformations between
(suitable open sets of) the degree $d_\CL$ Jacobian of the spectral curve
$R(\mu,\la)$ associated with $\CL(\la)$ and the (corresponding open subsets) 
of the Cartesian product of the  
degree $r(r-1)/2$ Jacobians associated with the curves
$\Gamma^a(\la,\mu)$\footnote{This observation is due to B. Dubrovin.}.
This simply follows from the fact that both the algebro-geometrical
coordinates found from $\CL(\la)$ and those we discovered in this paper are
{\em canonical coordinates} for the standard Lie-Poisson bracket $P_0$ 
on $sl(r)^n$, and, in particular, (together with those coordinates associated
with the global $SL(r)$ invariance) are Darboux coordinates for the restriction
of $P_0$ to its generic symplectic leaves.

It is outside the size of this paper to fully discuss this issue here. However
we think it is appropriate to display this transformation in the simple case
of $sl(2)$. This goes as follows.

We recall that the $n$ matrices $A_i$ can be explicitly 
parametrized by means of
$3n$ coordinates $h_i, e_i, f_i, \ i=1,\dots,n$:
\begin{displaymath}
A_i=\left[\begin{array}{cc} h_i& e_i\\ f_i&-h_i\end{array}\right],
\end{displaymath}
with Lie-Poisson brackets given by
\begin{displaymath}
\{ h_i, e_j \}_P= \delta_{ij} e_j,\quad
\{ h_i, f_j \}_P= -\delta_{ij} f_j,\quad
\{ e_i, f_j \}_P= 2 \delta_{ij} h_j
\end{displaymath}

Let us denote with $\lambda_i,\mu_i$ the 
set of separation variables associated with the Lax matrix 
$\CL(\la)$ (see eq. ~\rref{eq:6.2}) of the 
$n$-particles $sl(2)$-Gaudin model and with $\zeta_i,\rho_i$ those associated
with the \bih\ picture discussed in this paper.

We can assume that the coordinates associated with the global
$SL(2)$ invariance of the problem are the same; in particular, on the
symplectic leaves of the Lie Poisson tensor $P_0$ we have to consider
the pair 
\[ \lambda_1=\zeta_1=\sum_{j=1}^n f_j, \quad\mu_1=\rho_1= \frac{\sum_{j=1}^n h_j}{\sum_{j=1}^n f_j}
\]
According to the ``Sklyanin'' recipe, 
the $\lambda_a, \ a=2,\dots,n$ are the zeroes of
the $(1,2)$ entry of the rational Lax matrix (\ref{eq:1.2}), 
i.e. the roots  of the polynomial:
\begin{equation}
\Delta_\CL(\lambda)\equiv\sum_{k=0}^{n-1}\la^kC_k= \frac{1}{\sum_{i=1}^n f_i}
\sum_{k=0}^{n-1}\la^k \sum_{j=1}^{n}
(-1)^{n-k-1} s_{n-k-1}(a_1,\dots,\widehat{a_j}, \dots,a_n) f_j. 
\end{equation}
Here the polynomials $s_{n-k-1}(a_1,\dots,\hat{a_i}, \dots,a_n)$ are the
elementary symmetric polynomials in $n-1$ 
letters $b_1,\ldots, b_{n-1}$, defined by 
\begin{equation}\label{trasf11}
\prod_{j=1}^{n-1}(\la-b_j)=
\sum_{j=0}^{n-1}(-1)^{n-j-1} s_{n-j-1}(b_1, \dots, b_{n-1})\la^{j},\quad
\text{ where } s_0\equiv 1,
\end{equation}
evaluated for $b_1=a_1,\ldots,
b_{j-1}=a_{j-1},b_j=a_{j+1},\ldots,b_{n-1}=a_n$.

We can express the ``physical'' coordinates $f_i$ in terms of the $\la_i$ 
as follows:
\begin{equation}\label{trasf1}
\frac{f_i}{\sum_{j=1}^n f_j}= \frac{1}{\prod_{k \neq i} (a_i-a_k)}
\sum_{j=0}^{n-1} a_i^j C_j, \qquad
\sum_{j=1}^n f_j= \lambda_1.
\end{equation} 
Since the coordinates $\zeta_i$ are rational functions of the $f_i$ alone, and
namely,
\begin{displaymath}
\zeta_{a}= -\frac{ \sum_{k=1}^{a-1} f_k}{f_a}, \qquad a=2,\dots, n,
\end{displaymath}
we can explicitly find the transformation yielding the $\zeta_i$ in terms of
the $\la_i$ as:
\begin{equation} 
\zeta_{b}=-\sum_{k=1}^{b-1} \left( \frac{\Pi_{l \neq b} (a_{b}-a_l)}{\Pi_{l\neq k} (a_k-a_l)} 
\cdot\frac{\sum_{j=0}^{n-1} a_k^j (-1)^{n-j-1} s_{n-j-1}
(\lambda_2, \dots, \lambda_{n})}{\sum_{j=0}^{n-1} a_{b}^j
(-1)^{n-j-1} s_{n-j-1}(\lambda_2, \dots, \lambda_{n})} \right),
\label{trasfl}
\end{equation}
for $b=2,\ldots, n$, with, obviously 
\begin{equation}\label{trasfln}
\zeta_1=\lambda_1.
\end{equation}
The variables 
$\mu_a, a=2,\dots,n$ are the values
of the $(1,1)$ entry of the rational Lax matrix 
(\ref{eq:1.2}) for $\lambda=\lambda_a$,
while the $\rho_a$ are be given by the values of 
the $(1,1)$ entry of the
Lax matrix $L_{a}$ in $\lambda=\zeta_a$, 
divided by the normalizing factor $\zeta_a(\zeta_a-1)$.
Explicitly:
\begin{eqnarray*}
&& \mu_a=\sum_{j=1}^n \frac{h_j}{\lambda_a-a_j} \quad a=2,\dots,n, \qquad \mu_1=\frac{\sum_{j=1}^n h_j}{\lambda_1}\\
&& \rho_a=\frac{\zeta_a h_{a} + \sum_{j=1}^{a-1} h_j}{\zeta_a(\zeta_a-1)}  \quad
a=2,\dots,n,
\qquad\rho_1=\frac{\sum_{j=1}^n h_j}{\zeta_1}
\end{eqnarray*}
The transformation of coordinates connecting the $\rho_i$ and the $\mu_i$ 
can be easily found noticing that they are connected to the 
coordinates $h_i$ by a 
linear transformation with coefficients depending on $\zeta_i$ and
$\lambda_i$ respectively.
Consequently, the transformation between the coordinates $\rho_i$ and $\mu_i$ 
is a linear transformation with $\lambda_i$ depending coefficients:
\begin{equation}
\rho_i=\sum_{j=1}^n A_{ij}(\lambda_1,\dots,\lambda_n) \mu_j \label{linear}
\end{equation}
and it follows that it must be the lifting of the transformation defined by 
(\rref{trasfl} and \rref{trasfln})
 among the $\zeta_i$ and the $\lambda_i$;
\begin{equation}
\rho_i=\sum_{j=1}^n ((J^t)^{-1})_{ij} \mu_j \label{trasfm}
\end{equation}
where $J$ is the Jacobian of the transformation (\ref{trasfl}).

Finally, we just mention some other 
problems which remain open. The first one is to
compare our results with the picture of the generalized bending flows of
\cite{FlMi01}. The setting presented in that paper was aimed at providing a
generalization of the previous paper \cite{KM}, and, in our setting, should be
obtained by reduction to the submanifold of matrices $A_i$ having rank equal
to one. 

The second one is the application of the scheme herewith presented to the
quantum $sl(r)$ case. Preliminary results for $r=3$ indicate that this should
be a viable procedure for giving explicit expressions to the quantum integrals
whose existence has been proven in \cite{FFR}.  
Work in both these directions is in progress. 

\section*{Acknowledgments} We thank B. Dubrovin, J. Harnad, 
F. Magri, M. Pedroni, and O. Ragnisco 
for useful discussions. This work was partially
supported by GNFM-INdAM, and by the Italian MIUR under the project {\em
  Geometry of Integrable Systems}.
\newappendix{: Proof of Proposition \ref{prop:delta1}}
The proof of Proposition~\ref{prop:delta1} is divided into a couple of 
steps.
We recall that   we are considering 
a corank $k$ affine \bih\  manifold $(M, P_1-\la P_0)$, endowed with
a transversal distribution
$\CZ$, satisfying the requirements of Proposition \ref{prop:transdist}.
$Z_a$,
$\fino{a}{1}{k}$ is a set of normalized flat generators for $\CZ$, and
\[
\Ha{a}(\la)=\la^{n_a}H^a_0+\cdots+H^a_{n_a},
\]
are polynomial Casimirs of $P_1-\la P_0$. Finally,
the $k^2$
polynomials 
\begin{equation}
D^a_b(\la)=\Lie{Z_b}\Ha{a}(\la)=
\la^{n_a}\delta^a_b-D^a_{b,\, 1}\la^{n_a-1}-\cdots-D^a_{b,\,n_a}.
\end{equation}
are their deformations along the flat generators. Finally, we recall the 
definition
\[
\wid{P}_1=P_1-\sum_{a=1}^k X_a\wedge Z_a, \quad X_a=P_1 d H^a_0.
\]
\begin{lemma}\label{prop:bidiffideal}  
The actions of $\wid{P}_1$ and $P_0$ 
on the deformation of the Casimirs of the pencil are related by
the following formula:
\begin{equation}
  \label{eq:3.2.DNst}
  \wid{P}_1 d D^a_b(\la)=\la P_0 d D^a_b(\la)+\sum_{c=1}^k  D^a_c(\la) P_0 
dD^c_{b,\,1}.
\end{equation}
\end{lemma}
{\bf Proof.} We limit ourselves to sketch the proof of this Proposition, 
which is essentially contained in Section 7 of \cite{FP1}, although in a
disguised form. 
We consider the characteristic property of a Casimir of the
Poisson pencil,
\[
P_\la d \Ha{a}(\la)=0
\]
and derive it w.r.t. $Z_b$.  We get:
\begin{equation}
  \label{eq:3.2.zh}
  \Lie{Z_b}(P_\la)\, d\Ha{a}(\la)+P_\la d D^a_b(\la)=0.
\end{equation}
Since $\Lie{Z_b}(P_\la)=\sum_{c} [Z_b,X^1_c]\wedge Z_c$ with
$X^1_c=P_1 d H^c_0=P_0 dH^c_1$, we see that
\[
[Z_b,X^1_c]=\Lie{Z_b}(X^1_c)=\Lie{Z_b}(P_0dH^c_1)=-P_0 d D^c_{b,\,1}.
\]
Thus eq.\rref{eq:3.2.zh} takes the form
\begin{equation}
  \label{eq:3.2.zh2}
P_\la d D^a_b(\la)-\sum_c D^a_c(\la) P_0 d D^c_{b,\, 1}-\sum_c \langle
[Z_b,X^1_c],d\Ha{a}(\la)\rangle \cdot Z_c=0. 
\end{equation}
Let us consider the coefficient $\langle
[Z_b,X^1_c],d\Ha{a}(\la)\rangle$ in the last sum. This is, by definition,
\[
\Lie{ [Z_b,X^1_c]}
(\Ha{a}(\la))=\Lie{Z_b}\Lie{X^1_c}(\Ha{a}(\la))-\Lie{X^1_c}\Lie{Z_b}(\Ha{a}(\la)).
\]
Since $\{H^c_1,\Ha{a}(\la)\}_0=0$  only the second term is non identically
vanishing, and equals $-\Lie{X^1_c}(D^a_b(\la))$. 

Furthermore, thanks to the
affinity of the GZ manifold, we see that all terms of the form
$\Lie{Z_c}(D^a_b(\la))$ identically vanish.

So, we see that \ref{eq:3.2.zh2} can be written as
\begin{equation}
  \label{eq:3.2.zh3}
 P_\la d D^a_b(\la)-\sum_c D^a_c(\la) P_0 d D^c_{b,\, 1}-\sum_c 
(X^1_c\wedge Z_c)\cdot(d D^a_b(\la))=0, 
\end{equation}
which, in view of ~\rref{eq:3.op}, yields the statement.
\endpf

\begin{prop} \label{prop:charp}
Let $D$ be a $k\times k$ polynomial matrix of the form
  \begin{equation}
    \label{eq:3.2.polma}
    D^a_b(\la)=\la^{n_a}\delta_b^a-\la^{n_a-1}D^a_{b,\,1}-\cdots-D^a_{b,\,n_a},
\fino{a,b}{1}{k},
  \end{equation}
where the $D^a_{b,\,p}$ are smooth independent functions on 
a \varb\ $M$, satisfying
equation~\rref{eq:3.2.DNst}. Then:
\begin{enumerate}
\item
 Its determinant $\Delta(\la)$ has the form
 \begin{equation}
   \label{eq:3.2.de}
   \Delta(\la)=\la^\nu-\Delta_1\la^{\nu-1}+\cdots+\Delta_\nu, 
 \end{equation}
with $\nu=\sum_a n_a$ and satisfies
\begin{equation}\label{eq:3.2.det}
\wid{P}_1 d \Delta(\la)=\la P_0 d \Delta(\la)+ \Delta(\la) P_0 d\Delta_1.
\end{equation}
\item The roots $\la_i$ of $\Delta(\la)$ satisfy
\begin{equation}\label{eq:3.2.la}
\wid{P}_1 d \la_i=\la_i d\la_i.
\end{equation}
\item
Let $\wid{D}(\la)$ denote the classical adjoint matrix of ${D}(\la)$, and let
$[\wid{D}(\la)]_{a,c}$ be non identically vanishing. 
Then any  ratio
$\rho(\la):=[\wid{D}(\la)]_{a,b}/[\wid{D}(\la)]_{a,c}$ 
of elements belonging to the $a$-th row of $\wid{D}(\la)$,   
evaluated at the roots $\la_i$ of $\Delta(\la)$ satisfy the equation
\begin{equation}
  \label{eq:3.5nn}
  \wid{P}_1 d\rho(\la_i)=\la_i P_0d\rho(\la_i).
\end{equation}
\end{enumerate}
\end{prop}

{\bf Proof.} The power
expansion~\rref{eq:3.2.polma} simply states that the
$(a,a)$ entry of $D^a_b$ is a monic
degree $n_a$ polynomial, while all other entries in the $a$-th row are of
degree not exceeding $n_a-1$. We preliminarily notice that
\begin{equation}\label{eq:3.2.mn}
\Delta(\la)=\prod_{a=1}^k D^a_a(\la)+O(\la^{\nu-2}),\> \text{whence }
\Delta_1=\sum_{a=1}^k D^a_{a,\,1}.
\end{equation}
We multiply the matrix 
equation~\rref{eq:3.2.DNst} say, on the left, by 
the classical adjoint $\widetilde{D}$, to get
\begin{equation}
  \label{eq:3.2.adj}\begin{split}
  \sum_c \wid{P}_1&({\widetilde{D}}^a_c d(\la) D^c_b(\la))-  \sum_c \la P_0
{\widetilde{D}(\la)}^a_c d D^c_b(\la)=\\ 
&\sum_{c,d}  
\widetilde{D}^a_c(\la) D^c_d (\la)P_0 d D^d_{b,\,1}.\end{split}
\end{equation}
Recalling that $\sum_{c}  
\widetilde{D}^a_c (\la)D^c_d (\la)=\delta_{a d}\Delta(\la)$ and 
\[
\sum_{a,c}\widetilde{D}^a_c (\la)d D^c_a (\la)= 
\text{Tr}\widetilde{D}(\la) dD=d\Delta(\la),
\]
taking the trace of the matrix equation \ref{eq:3.2.adj} and taking into
account~\rref{eq:3.2.mn} we get the proof of the first item.

To prove item \# 2, we first notice that, for any function $f$ on $M$, the
evaluation of a polynomial (or rational) function $F(\la)$ 
in the parameter $\la$,
whose coefficients are themselves functions on $M$ gives rise to a new
function $F(f)$ on $M$. Its differential can be written as follows:
\[
d(F(f))=dF(\la)\Big \vert_{\la=f}+\ddd{F(\la)}{\la}
\Big\vert_{\la=f} df
\]
To clarify the notations, the first term in the RHS of the above equations
means the differential of $F(\la)$, with $\la$ taken as a parameter, then
evaluated for $\la=f$, and the second the partial derivative of $F(\la)$
w.r.t. $\la$, subsequently evaluated for $\la=f$.
Keeping this proviso in mind, we consider now $F(\la)=\Delta(\la)$, and
$f=\la_i, \fino{i}{1}{n}$. We have: 
\begin{equation}
  \label{eq:3.2.diff}
  d(\Delta(\la_i))=d\Delta(\la)\Big\vert_{\la=\la_i}+\ddd{\Delta(\la)}{\la}
\Big\vert_{\la=\la_i} d\la_i,
\end{equation}
where $d\Delta(\la)=-\sum_j \la^{p-j} d\Delta_j $. Taking into account the
relation \rref{eq:3.2.diff}, we get
\begin{equation}
  \label{eq:3.2.dif2}
  0=(\wid{P}_1-\la_i P_0)d\Delta(\la_i)=
\Delta(\la_i) P_0 d\Delta_1+\ddd{\Delta(\la)}{\la}
\Big\vert_{\la=\la_i} (\wid{P}_1-\la_i P_0)d\la_i,
\end{equation}
which implies the assertion, since
\[
\Delta(\la_i)=0,\> \text{while } \ddd{\Delta(\la)}{\la}
\Big\vert_{\la=\la_i}\neq 0,
\]
thanks to the fact that being the coefficients $\Delta_{a,i}$ functionally
independent, the roots are generically simple.

The proof of the third assertion is basically contained in  the
proof of Proposition 8.4 of \cite{FP1}. We limit ourselves to sketch it.

By using the relations~\rref{eq:3.2.DNst} and \rref{eq:3.2.det}, 
together with the defining relation 
$\widetilde{D}^a_c (\la)D^c_d(\la)=\delta_{a d}\Delta(\la)$, one
arrives at the matrix equation
\begin{equation}
  \label{eq:3.2.nu}
  \sum_{c}\big(\wid{P}_1 d \wid{D}^a_c -\la P_0 d \wid{D}^a_c\big)
  D^c_b=\Delta(\la)P_0\big(d \Delta_1 \delta_{b}^a-d D_{b,1}^a\big).
\end{equation}
If $\boldsymbol{\sigma}$ denotes one row of the adjoint matrix 
$\wid{D}(\la)$, we can rewrite the above equation as
\[
\big(\wid{P}_1\boldsymbol{\sigma}  -\la P_0\boldsymbol{\sigma}
\big)D(\la)=\Delta(\la)\cdot X
\]
where $X$ is the corresponding row of the RHS of ~\rref{eq:3.2.nu}.
If we consider the normalized row
$\boldsymbol{\rho}=\boldsymbol{\sigma}/\sigma_j$, we see that, since
\[
\boldsymbol{\rho} D(\la)=\Delta(\la)\boldsymbol{\sigma}/\sigma_j,
\]
it holds
\begin{equation}\label{eq:3.2.u}
\big(\wid{P}_1\boldsymbol{\rho}  -\la P_0\boldsymbol{\rho}
\big)D=\Delta(\la)\cdot Y,
\end{equation}
for some suitable $Y$ whose form is irrelevant here.
Evaluating this equation for $\la=\la_i$, we see that
\begin{equation}\label{eq:3.2.u2}
\big(\wid{P}_1\boldsymbol{\rho}  -\la P_0\boldsymbol{\rho}
\big)D(\la)\Big\vert_{\la=\la_i}=0
\end{equation}

Taking  into account that $\Delta$ has simple eigenvalues, we see
that each row $(\wid{P}_1 d\boldsymbol{\rho}  -\la P_0 d\boldsymbol{\rho})$, 
evaluated at $\la=\la_i$ must be
proportional to the corresponding row of $\wid{D}(\la)$, 
that is, there must exist vector fields $X'$  such that
\[
(\wid{P}_1 d \boldsymbol{\rho} -\la P_0 d \boldsymbol{\rho})\vert_{\la=\la_i}
=X'\cdot \boldsymbol{\rho}\Big{\vert_{\la=\la_i}}.
\] 
Since one element of $\boldsymbol{\rho}$ is normalized to $1$, we thus see
that $X'$ must vanish, whence the thesis. 
\endpf
\newappendix{: Proof of Proposition \ref{prop:b2}}
The key point is the following
observation on the (ordinary) Lie-Poisson brackets on a single copy of
$M=sl(r)$. The Poisson bracket of two functions $F,G$ on $M$,
is given by
$\{F,G\}={\rm Tr}(\ddd{F}{A}\cdot [A,\ddd{G}{A}])=
-{\rm Tr}(A\cdot[\ddd{F}{A},\ddd{G}{A}]).$
Let $A_{ij}$ denote the $ij$-th entry of $A$, and consider the family of
$r-1$ vector fields on $M$ defined by $W^\al=\ddd{}{A_{r,\al}},
\> \al=1,\ldots, r-1,$
as well as the distribution $\CZ \subset TM$ defined by the
$W^\al$.

We notice that differentials of
functions vanishing along $\CZ$ admit a very simple matrix
representation. Indeed $W^\al$ is represented via its action on the 
matrix $A$ as the elementary matrix $e_{\al,r}$, having $1$ in the $\al$-th
place of the last column.
So $\Lie{W^\al}(F)=0$ iff $\left(\ddd{F}{A}\right)_{r,a}=0, a=1,\ldots,r-1$, 
i.e., iff $\ddd{F}{A}$
lies in the lower maximal parabolic subalgebra $\alp$ of $sl(r)$.

Let now $W$ denote any element in $\CZ$, and let
$F,G$ be functions such that $\Lie{Z}F=\Lie{Z}G=0$, and
let us compute $\Lie{Z}(\{F,G\})$. Thanks to the Leibniz property of
the Lie derivative and the fact
that $Z$ is a constant vector field we have that
\begin{equation}\label{eq:zff0}
\Lie{W}(\{F,G\})
=-{\rm Tr} \left( \Lie{W}(A)\cdot \left[ \ddd{F}{A},\ddd{G}{A} \right] 
\right)
\end{equation}
which vanishes as well since $\alp$ is indeed a Lie subalgebra of $sl(r)$.

In the case  of the $n$-particle $sl(r)$ Gaudin 
model, whose phase space is parametrized by $n$ matrices $A_i$,
we consider the family of $n\cdot (r-1)$ vector fields 
defined by
\begin{equation}
\Lie{ W_i^\al}(A_1,\ldots, A_n)=(0,0,\cdots,\underbrace{e_{\al,r}}_{\text{i-th
    place}},\cdots,0).
\end{equation}
The distribution $\CZ$ generated by these vector fields is generically
transversal to the symplectic leaves of the Lie--Poisson product structure
on $sl(r)^N$. 
We now prove that the space of
functions vanishing along $\CZ$ is a Poisson subalgebra for any {\em{affine}} 
Poisson tensor $Q$. 
The brackets $\{F,G\}_Q=\langle dF,Q dG \rangle$ are given by the multiple sum
\[
\{F,G\}_Q=\sum_{i,j,k=1}^N{\rm Tr}\left(\ddd{F}{A_i}\cdot\left(\sum_{k=1}^N 
c^k_{i,j}\left[ A_k,\ddd{G}{A_j} \right] \right)+d^k_{i,j}
\left[\sigma_k,\ddd{G}{A_j} \right]\right)
\]
where $\sigma_k$ denote constant matrices.
Noticing that the differentials of functions $F$ vanishing along 
$\CZ$ are represented by
$n$-tuples of matrices $dF=(\ddd{F}{A_1},\ddd{F}{A_2},\ldots,\ddd{F}{A_n})$ 
with $\ddd{F}{A_i}\in \alp, \> i=1,\ldots,N$, 
we see that the Lie derivatives $\Lie{W_i^\al}\{F,G\}_Q$ are given by 
multiple sums of terms like those of
Eq.~\rref{eq:zff0}, and so vanish whenever $\Lie{W_i^\al}F=\Lie{W_i^\al}
G=0$. 
\endpf

\newappendix{: Proof of Proposition \ref{prop:penult}}
The proof of Proposition \ref{prop:penult} follows
from a few elementary but important facts following 
from the definitions
of the Casimirs of the Poisson pencil $\Ka{i}(\la)$ and of the normalized
transversal vector fields $Z^\al_i,\al=1,\ldots,r-1,\> i=1,\ldots,n$.
Recall that  we defined $n^2$ matrices $D_{ij}$  by
\begin{equation}
\big(D_{ij}\big)^\al_\be=\Lie{Z^\be_j}(\Ka{i}_\al(\la)),
\end{equation} 
where the polynomial Casimirs $\Ka{i}_\al$ are defined by 
\begin{equation}
  \label{eq:4.1.1.1}
  \text{Det}(M_i(\la,\xi))=\xi^r+\sum_{\al=1}^{r-1}\Ka{i}_\al(\la) 
\xi^{r-\al-1}.
\end{equation}

One has:
\begin{enumerate}
\item The matrix $\big(D_{11}\big)$ is the identity. This trivially follows
  form our choice of $L_1=A_1$.
\item The $n(r-1)\times n(r-1)$ matrix of the deformations of the
  Casimirs w.r.t. the transversal vector fields has the following block form:
  \begin{equation}
    \label{eq:4.1.blofo}
    {\DD}=\left[\begin{array}{ccccc}
D_{11}&0&\cdots&\cdots& 0\\
D_{21}&D_{22}&0&\cdots&0\\
\vdots&&\ddots&\vdots&0\\
D_{n1}&D_{k3}&\cdots&&D_{nn}
\end{array}\right]
  \end{equation}
This follows form the fact that, for $j>i$ and every $\al$, 
$\Lie{Z^\al_j} L_{j}(\la)=0$.
Thanks to point 1 above, we now consider the non trivial matrices 
$M_a(\la,\xi), \finon{a}{2}$, and the corresponding fields $Z^\al_b$.

Taking into account that:
\begin{description}
\item[i)]
$
\Lie{Z^\al_b}(\text{Det}(M_a(\la,\xi)))=-\text{Tr}(\widetilde{M}_a(\la,\xi)
\Lie{Z^\al_b}(L_a))$;
\item[ii)]
$\Lie{Z^\al_a}(L_a)= (\la-a+2)\Lie{Z^\al_a}(A_a)$;
\item[iii)]
The determinants of the diagonal blocks
$\text{Det} D_{aa}=\CD^a(\la)$ 
are monic polynomials of degree $r(r+1)/2-1$ in
$\la$;
\end{description}
We can factorize $\CD^a(\la)$  as
\begin{equation}\label{eq:4.1.det}
\CD^a(\la)=(\la-a+2)^{r-1}\Delta^a(\la)
\end{equation}
where $\Delta^a(\la)$ is a monic polynomial of degree $r(r-1)/2$.
\item Thanks to the lower diagonal block form of the matrix $\DD$ of
  equation~\rref{eq:4.1.blofo}, {\em every}  diagonal block $D_{aa}$ satisfies 
Proposition~\rref{prop:bidiffideal}. So its determinant satisfy, according 
to Proposition \ref{prop:charp}
$  (\wid{R}-\la P) d \CD^a(\la) = \CD^a(\la)  P d \CD^a_{1},$
and thanks to the factorization property \rref{eq:4.1.det}, 
we have
\begin{equation}
  \label{eq:4.1.detred}
  (\wid{R}-\la P) d {\Delta^a(\la)}=\Delta^a(\la) P d \Delta^a_{1}.
\end{equation}
\end{enumerate}

We now recall that the Casimir functions of $P$ are given by the highest order
terms $\Ka{i}_{\al,0}$ of the expansion of the Casimirs of the Poisson pencil
$\Ka{i}_\al(\la)$ in powers of $\la$. If we call 
\begin{equation}\label{eq:normaliz}
G^{\al,\be}_i=\Lie{W^\al_i}\Ka{i}_{\beta,0}
\end{equation}
the matrix of the deformations of the Casimirs of $P$ with respect to the
corresponding vector fields $W^\al_i$ introduced in eq.~\rref{eq:b4}, noticing
that $
\Lie{W^\al_i}\Ka{j}_{\beta,0}$ vanishes for $j\neq i$,
we see that the normalized
generators $Z^\al_i$ and the ``constant'' ones $W^\al_i$ are related by 
\[
W^\al_i=\sum_\be G^{\al,\be}_i Z^\be_i
\>\finon{i}{1}.
\]
Thus, considering only the nontrivial indexes $\finon{a}{2}$,
\begin{equation}
  \label{eq:4.1.emat}
  \Lie{W^\al_a}\text{Det}(M_a(\xi,\la))=\sum_\be G^{\al,\be}_a
  \Lie{Z^\al_a}\text{Det}(M_a(\xi,\la)),
\end{equation}
we can argue as follows.
Since $G^{\al,\be}_a$ are independent of $\la$, we see that the common
solutions $(\xi_a,\la_a)$ of the two sets of $r-1$-tuple of equations
\begin{equation}\label{eq:fghw}
\left\{\begin{array}{c}
\Lie{W^2_a}\text{Det}(M_a(\xi,\la))\\ \vdots\\
\Lie{W^r_a}\text{Det}(M_a(\xi,\la))\end{array}\right.
\end{equation}
and 
\begin{equation}\label{eq:fghz}
\left\{\begin{array}{l}
\Lie{Z^2_a}\text{Det}(M_a(\xi,\la))\\ \vdots\\
\Lie{Z^r_a}\text{Det}(M_a(\xi,\la))\end{array}\right.
\end{equation}
coincide for every (fixed) $\finon{a}{2}$.

Expanding the RHS of the equations of the system \rref{eq:fghz} as
\begin{equation}\label{eq:solita}
\Lie{Z^\al_a}\text{Det}(M_a(\xi,\la))=\sum_{\be=1}^{r-1} 
\xi^{r-1-\be}\Lie{Z^\al_a}\Ka{a}_\beta=\sum_{\be=1}^{r-1} 
\xi^{r-\be}(D_a)^{\al}_{\be},
\end{equation}
we see that:
\begin{description}
\item[a)] The roots of $\CD^a(\la)$, introduced in eq. \rref{eq:4.1.det} 
are those values of $\la$ for which the $r-1$
  equations, defined for $\finon{a}{2}$,
  \begin{equation}
    \label{eq:4.1.dudi}
    \sum_{\be=1}^{r-1}\xi^{r-1-\al}(D_a)^\al_\be=0
  \end{equation}
admit solutions. In particular, the
roots of $\Delta^a(\la)$ define non trivial elements
$\la_a^s,\fino{s}{1}{r(r-1)/2}$.
\item[b)] The values $\xi_a^s$ corresponding to the roots $\la_a^s$  of
  $\Delta^a(\la)$ 
are given by suitably normalized elements of the adjoint matrix 
$\widetilde{D}_{a}$, evaluated at $\la=\la_a^s$. 
\end{description}
Thus, from the \bih\ theory, we can conclude that the only non vanishing 
Poisson brackets between such functions admit the separate form:
\begin{equation}
  \label{eq:4.nnn.pb}
  \{\la_b^s,\xi_a^t\}_{P}=\delta^{st}\delta_{ab} 
\varphi^{s}_{a}(\xi_a^s,\la_a^s), \qquad \{\la_b^s,\xi_a^t\}_\wid{R}=
\delta^{st}\delta_{ab} \la_a^s\varphi^{s}_{a}(\xi_a^s,\la_a^s)
\end{equation}

We now consider eq. \rref{eq:fghw}, taking into account the observation that
the pairs $(\la_b^s,\xi_a^t)$ are solutions of this system as well. We notice
that from the definition of $W^\al_a$, 
$\Lie{W^\al_a}\text{Det}(M_a(\xi,\la))$ is nothing but the determinant of
the minor of $M_a(\xi,\la)$ relative to the $\al,r$ entry.
Since the $r(r-1)/2$ pairs $(\xi^a_i,\la^a_i)$, for every fixed $a$,
  annihilate the $r-1$ minors of the matrix $M_a(\xi,\la)$ relative to the 
entries $(1,r),\ldots(r-1,r)$, they annihilate the minor relative to the
$(r,r)$ entry as well. Hence they annihilate the last row of the adjoint
matrix of $M_a(\xi,\la)$, and so satisfy the characteristic equation
\begin{displaymath}
\text{Det}\big(M_a(\xi_a^s,\la_a^s)\big)=0,
\quad \finon{a}{2},\> \fino{s}{1}{r(r-1)/2}. 
\end{displaymath}

\end{document}